\documentclass[sigconf, nonacm, authorversion]{acmart}

\usepackage{dirtytalk} 
\usepackage{wrapfig}

\usepackage{subcaption}

\usepackage[bottom]{footmisc}

\definecolor{myRed}{RGB}{255, 45, 85}
\definecolor{myGreen}{RGB}{76, 217, 100}
\definecolor{myBlue}{RGB}{0, 122, 255}
\definecolor{myOrange}{RGB}{255, 136, 0}
\definecolor{darkergreen}{RGB}{37, 155, 57}
\definecolor{myTeal}{RGB}{181, 255, 251}
\definecolor{myLightGray}{RGB}{213, 217, 224}

\usepackage{xcolor}
\usepackage{enumitem}
\usepackage{soul,xcolor}

\newlist{steps}{enumerate}{1}
\setlist[steps, 1]{label = Step \arabic*:}

\AtBeginDocument{%
  \providecommand\BibTeX{{%
    \normalfont B\kern-0.5em{\scshape i\kern-0.25em b}\kern-0.8em\TeX}}}

\setcopyright{acmcopyright}
\copyrightyear{2023}
\acmYear{2023}
\acmDOI{XXXXXXX.XXXXXXX}

\acmPrice{15.00}
\acmISBN{978-1-4503-XXXX-X/23/10}

\begin{document}

\title{Promptify: Text-to-Image Generation through Interactive Prompt Exploration with Large Language Models}




\author{Stephen Brade}
\affiliation{%
  \institution{University of Toronto}
  \city{Toronto}
  \country{Canada}
}
\email{stephen.brade@mail.utoronto.ca }

\author{Bryan Wang}
\affiliation{%
  \institution{University of Toronto}
  \city{Toronto}
  \country{Canada}
}
\email{bryanw@dgp.toronto.edu}

\author{Mauricio Sousa}
\affiliation{%
  \institution{University of Toronto}
  \city{Toronto}
  \country{Canada}
}
\email{mauricio@dgp.toronto.edu}

\author{Sageev Oore}
\affiliation{%
  \institution{Dalhousie University}
  \city{Halifax}
  \country{Canada}
}
\email{sageev@dal.ca}

\author{Tovi Grossman}
\affiliation{%
  \institution{University of Toronto}
  \city{Toronto}
  \country{Canada}
}
\email{tovi@dgp.toronto.edu}

\renewcommand{\shortauthors}{Brade et al.}

\begin{abstract}
Text-to-image generative models have demonstrated remarkable capabilities in generating high-quality images based on textual prompts. However, crafting prompts that accurately capture the user's creative intent remains challenging. It often involves laborious trial-and-error procedures to ensure that the model interprets the prompts in alignment with the user's intention. To address these challenges, we present Promptify, an interactive system that supports prompt exploration and refinement for text-to-image generative models. Promptify utilizes a suggestion engine powered by large language models to help users quickly explore and craft diverse prompts. Our interface allows users to organize the generated images flexibly, and based on their preferences, Promptify suggests potential changes to the original prompt. This feedback loop enables users to iteratively refine their prompts and enhance desired features while avoiding unwanted ones. Our user study shows that Promptify effectively facilitates the text-to-image workflow, allowing users to create visually appealing images on their first attempt while requiring significantly less cognitive load than a widely-used baseline tool.
\end{abstract}







  \begin{teaserfigure}
    \centering
    \includegraphics[width=0.95\textwidth]{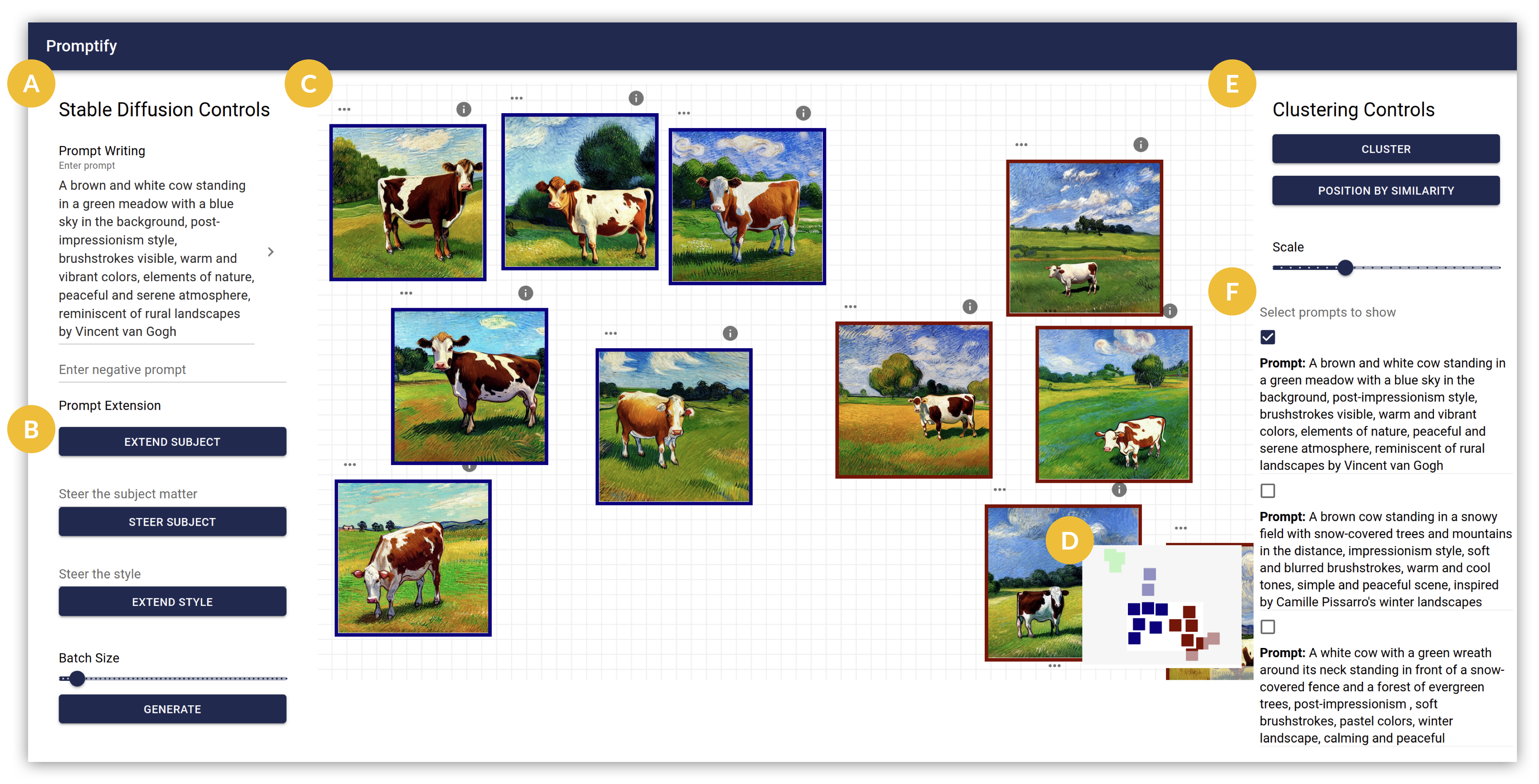}
    \caption{
    The Promptify System UI. 
    A) Stable Diffusion Controls for prompt writing. 
    B) Automatic Prompt Suggestion for ideating subject matter and obtaining keywords suggestions for style in a steerable manner via natural language. 
    C) Image Layout and Clustering allows users to view, organize, and cluster their generated images. Users can interact with images and clusters by dragging, dropping, and zooming. They can also access prompt refinement suggestions based on individual images or clusters. 
    D) Minimap provides visual cues for clusters and helps users navigate the visualization.
    E) Clustering Controls for positioning and clustering images by similarity and adjusting the spacing between images via the "Scale" slider. 
    F) Prompt History allows users to toggle on/off all previous prompts to view or remote images using the prompt history menu.
    }
    \Description{Screenshot of Promptify user interface web page. The web page consists of the large central visualization or the layout and clustering feature showing generated images. On the left are the areas for writing prompts and ideating subject matter and extending the style of a user's prompts. On the right are the controls for the for the central layout and clustering feature.}
    \label{fig:teaser}
  \end{teaserfigure}

\maketitle

\section{Introduction}
Text-to-image generative models such as Stable Diffusion (SD)~\cite{rombach2021highresolution} and DALL-E-2~\cite{ramesh2022hierarchical} are capable of producing high-quality images based on natural language descriptions. 
These models can support creative processes in various fields, such as creating news illustrations~\cite{liu2022opal} and generating ideas for industrial design~\cite{Ko2022LargescaleTG}. 
However, recent studies have revealed that crafting prompts--the primary means of steering image generation--can be a challenging task~\cite{Ko2022LargescaleTG, liu2022opal, maneesh2023}. 
Crafting prompts is difficult because users need to learn how to create prompts that models can interpret in a manner that aligns with their intended creative output. 
This often involves an iterative trial-and-error process, where users write a prompt, observe the generated output, and refine their prompts if necessary. 
Compounding the difficulty of this process is the current models' lack of built-in support for discovering useful keywords that can aid in improving output quality and exploring alternative possibilities.

 Previous prompt engineering studies addressed these challenges by offering effective strategies for prompt writing~\cite{liu2022promptenineering, oppenlaender2022taxonomy, Saravia_Prompt_Engineering_Guide_2022, witteveen2022investigating, liu2021makes, wang2023reprompt}. 
 For example, Liu and Chilton~\cite{liu2022promptenineering} proposed to prioritize the identification of keywords for subjects and styles, and found that varied phrasings of the same keywords do not result in substantial differences. 
 Moreover, online communities such as the subreddit \texttt{r/StableDiffusion}\footnote{r/StableDiffusion: \url{https://www.reddit.com/r/StableDiffusion/}} have been formed to share prompt examples and writing strategies. 
 
 While these studies provide valuable insights, they only offer high-level advice and cannot make personalized recommendations for users seeking specific aesthetics. 
 Consequently, novice users lacking prior experience in prompt writing and familiarity with relevant keywords may still face difficulties achieving their desired results. To better understand the practices and challenges of prompting text-to-image models, we conducted formative user interviews with six active members of the online community. Participants highlighted several commonalities, including the importance of community resources for learning how to prompt, the iterative nature of prompt refinement, and the need for generating large image collections for inspiration and exploration.

 In this paper, we present Promptify, an interactive system to facilitate iterative prompt exploration and refinement for SD~\cite{rombach2021highresolution}, a popular open-source text-to-image diffusion model. Promptify facilitates the iterative workflow by supporting various steps, including ideation of subject matters, writing style modifiers, generating and browsing images, and refining the original prompt (Figure \ref{fig:teaser}).

To use our system, users start by inputting a subject, and Promptify suggests different ideas to extend the subject. They can further steer the subject suggestions to explore alternative ideas. Next, users input a short style description; Promptify then extends relevant modifiers 
\footnote{Keywords and modifiers for prompts are used interchangeably in the paper.} 
to articulate the style. The prompt suggestion engine can generate sophisticated prompts from users' brief input, leading to visually appealing images. Promptify visualizes the generated images on an image layout and clustering interface, allowing users to flexibly organize and browse the generated collections. Lastly, Promptify suggests keywords based on generated images to assist users in refining the original prompts to enhance desired features and avoid unwanted ones.

We developed Promptify's prompt suggestion engine using a chat-optimized large language model (LLM) GPT-3.5 ~\cite{openaiapi}, allowing users to iteratively improve the subject matter suggestions using natural language instruction. For style descriptions, we leverage few-shot prompting techniques to guide the LLM in making keyword suggestions resembling the language used by the community. 

To evaluate the effectiveness of Promptify, we conducted a 14-participant user study comparing Promptify with Automatic1111 ~\cite{automatic1111}, a popular tool for SD widely used by the community. Participants found Promptify significantly more useful in assisting the text-to-image generation workflow than the baseline tool and were able to create more visually appealing images on their first attempt. Promptify also assisted users in managing and comparing a large number of images to identify and reinforce desired features while ignoring unwanted ones in future iterations. Moreover, users reported experiencing significantly less mental demand and frustration when using Promptify. Taken together, our work makes the following contributions:
\begin{itemize}
    \item The design and implementation of Promptify, a full-stack system supporting interactive prompt exploration and refinement to facilitate text-to-image generation for novice users. 
    \item An LLM-based prompt suggestion engine that assists users in ideating subject matters and suggests style descriptions in text prompts based on community data.
    \item An interface for layout and clustering of generated images by similarity, enabling users to discover thematic trends and refine prompts for improving future generations.
    \item The results of a 14-participant user study demonstrating the efficacy of the Promptify system and its advantages over a commonly used baseline tool.
    
\end{itemize}
\section{Related Work}

\subsection{Text-to-Image Generative Models} 

Text-to-Image generative models refer to a class of machine learning models that generate images matching a language description as input~\cite{zhang2017stackgan, nguyen2017plug, mansimov2016generating, rombach2021highresolution, ramesh2022hierarchical, pmlr-v48-reed16, ramesh2021zeroshot}. 
One of the earliest models, AlignDraw~\cite{mansimov2016generating} learned to estimate alignment between text and the generating canvas. 
Reed et al. used convolutional generative adversarial networks (GANs) to generate plausible 64×64~\cite{pmlr-v48-reed16} and 128x128~\cite{reed2016learning} images for birds and flowers based on text descriptions. 
StackGAN~\cite{zhang2017stackgan} further improved the generation resolution to 256×256 by stacking two GANs, one for generating the primitive shape and basic colors, and the other for super-resolution. 
In addition to GAN-based models, OpenAI developed DALL-E~\cite{ramesh2021zeroshot}, a transformer-based model that was among the earliest text-to-image models to gain significant public attention. 
Its successor, DALL-E-2~\cite{ramesh2022hierarchical} uses CLIP (Contrastive Language-Image Pre-Training) embedding~\cite{radford2021learning} and a diffusion model~\cite{pmlr-v37-sohl-dickstein15} to produce more intricate images with 4x greater resolution. 
SD~\cite{rombach2021highresolution} uses a latent diffusion model to generate 512x512 images and is open-sourced to the public. Promptify uses SD as the basis due to its widespread popularity and availability.


\subsection{Prompt Engineering}

There has been a surge in generative models that use text as input to produce output, including autoregressive LLMs like the GPT series~\cite{brown2020language, openai2023gpt4}, and text-to-image models like SD~\cite{rombach2021highresolution}. 
The input text, known as a prompt, guides the model on what specific output to generate. Models like SD also utilize a negative prompt which steers the model away from undesirable attributes.
Prompt Engineering~\cite{liu2022promptenineering, oppenlaender2022taxonomy, Saravia_Prompt_Engineering_Guide_2022, witteveen2022investigating, liu2021makes, wang2023reprompt, hardprompts} refers to methods for constructing prompts to effectively communicate with models and direct the generation toward desired outcomes. 
The methods vary across different model modalities, e.g., text and images. 
For instance, few-shot prompting~\cite{brown2020language, liu2021makes, zhao2021calibrate} is a notable prompting technique for LLMs. 
This prompting technique involves providing the model with a small set of data examples from a specific task, such as translation, to guide the model in learning how to perform the task (in-context learning).
As for text-to-image models, studies have focused on identifying keywords or key phrases~\cite{oppenlaender2022taxonomy} and testing how various phrasings may result in different outputs~\cite{liu2022promptenineering}. 
3DALL-E~\cite{liu20223dalle} uses GPT-3 to suggest keywords for generating 3D designs. 
Wang et al.~\cite{wang2023reprompt} proposed an automatic method to refine text prompts toward precise emotional expressions of the generated images. Prompt-to-Prompt ~\cite{hertz2022prompt} supports image editing by editing an initial prompt. CLIPInterrogator~\cite{clipinterrogator} is an open-source tool that attempts to reverse-engineer an image by generating text prompts that are likely to match a given image. Promptify builds upon prior research and introduces a novel prompting method that establishes a connection between prompt engineering techniques for both LLMs and text-to-image models. We leverage few-shot prompting to direct LLMs in facilitating prompt engineering for SD models. Promptify also integrates CLIPInterrogator \cite{clipinterrogator} to assist users in identifying potential keywords that may enhance their results.

\subsection{Design Exploration and Semantic Editing}
A significant body of research in graphics and HCI has concentrated on aiding the generation, management, and exploration of design variations \cite{design_galleries, side_views, dream_lens, gemni, attribit, semantic_shape, procedural_modelling, intuitive_control}. Design Galleries \cite{design_galleries} utilizes visualizations to assist users in organizing the outcomes of diverse image rendering parameters. Dream Lens \cite{dream_lens} presents a visualization where users can overlay generated designs and remove undesired elements using a chisel tool. GEM-NI \cite{gemni} is a generative design tool with a graph-based interface that enables users to explore creative alternatives simultaneously.  Other work has focused specifically on creating semantic controls for visual content generation \cite{attribit, semantic_shape} or providing intuitive control of high dimensional design spaces with dimensionality reduction techniques \cite{procedural_modelling, intuitive_control}. Promptify contributes to this area by offering tools supporting users to explore variations in the domain of text-to-image generation.

\subsection{User Interfaces for Image Generation}

As generative models become capable of rendering high-quality images, there is an increasing focus on designing interfaces to deliver the power of these models to users~\cite{Evirgen_2022, evirgen2023ganravel, liu2022opal, liu20223dalle}. 
For instance, GANzilla~\cite{Evirgen_2022} enables users to iteratively explore various directions and achieve their editing goals using the scatter/gather technique. 
In the authors' subsequent study, they presented GANravel~\cite{evirgen2023ganravel} which proposes a user-driven direction disentanglement tool allowing users to enhance their editing directions iteratively. 
Opal~\cite{liu2022opal} is an interactive system designed to aid in creating news illustrations using text-to-image generation. 
3DALL-E~\cite{liu20223dalle} translates a designer's intentions into multimodal (text and image) prompts, which can generate 3D design inspirations. 
Contributions outside of academic research include open-sourced tools, such as GANInterface~\cite{GANInterface}, Easy Diffusion~\cite{EasyDiffusion}, and AUTOMATIC1111's Stable Diffusion WebUI~\cite{automatic1111}, which we refer to as Automatic1111 in the rest of paper. 
Promptify adds to this research area by contributing a novel interface that positions and clusters images by CLIP embedding similarity, providing a more efficient browsing experience than the traditional folder view.

\section{Formative Interviews}
To understand current practices and limitations of how users of SD generate desirable images, we interviewed six Reddit users who had posted multiple times on \texttt{r/StableDiffusion}. Particularly, we were interested in their experiences when they just began using SD, which is typically when users encounter the greatest challenges. The interviews were conducted via half-hour-long video conference calls, and the participants were compensated with 20 CAD. The participants had diverse levels of experience. Two of them self-reported as professional creators of digital art (P1, P4), two are hobbyists (P2, P3), and two are novices (P5, P6). All participants reported utilizing Automatic1111 on a personal GPU.The participants explained that the popularity of Automatic1111 was due to its effortless installation and support for a broad range of open-source tools, which were utilized to improve and manipulate the generated images. Below we discuss common challenges and the strategies participants have employed to overcome them.

\subsection{Prompt Writing is Difficult for Beginners}
Participants with varied experience levels reported significant difficulties writing effective prompts as beginners. P2, who creates visuals for role-playing campaigns, noted that it took them three months to produce something they were satisfied with. Additionally, transitioning to a new campaign may require several evenings of prompt experimentation to achieve satisfactory outcomes. P4 made a similar note about generating images for new subject matters and styles, remarking that \textit{"For every kind of output, my process becomes different."} P2 and P5 also noted that it is beneficial to understand the training data used for SD as it provides an intuition for which keywords will work. This, however, renders barriers for users without a technical background. 

The limited discoverability of effective prompting keywords compelled users to seek guidance from online communities. P6 noted \textit{"I was following SD on Reddit since it came out and tried to mimic what they were doing there"}. P5 mentioned that users on Reddit \textit{"put in a lot of effort to document what is working and what is not"} by sharing prompts and other hyperparameters. P4 mentions learning from Youtube videos and Discord forums and switching to SD from Midjourney ~\cite{Midjourney} \textit{"because SD is open source and people were sharing their prompts"} while P2 \textit{"[reaches] out on Discord to ask for help"} when they are not satisfied with a prompt. This demonstrates that participants improved their abilities by examining and emulating the prompts posted by others or by searching for similar prompts when their own did not yield satisfactory results. Some tools allow users to retrieve, e.g., \textit{lexica.art} \cite{lexica}, or generate prompts, e.g., \textit{MagicPrompt}~\cite{magicprompt}.  However, P3 commented that these models tend to produce prompts of a particular style, and they had to use regular expressions to filter out prompts containing keywords associated with their target styles, which is time-consuming.

\subsection{Iterative Experimentation is Required}
Participants (P1, P2, P6) reported that achieving satisfactory results often requires an iterative process of writing and refining a prompt, particularly when they seek to deviate from the community's stylistic trends. For example, P2 maintains a repository of effective keywords. To expand this collection, P2 generates two sets of images: one with prompts featuring a new keyword and the other without to determine the impact of each on the generated output. Other participants describe their iterative process of refining prompts to reinforce preferred outcomes and discourage undesired ones. For example, P1 notes that when they intended their subject to have pink hair, but SD created a subject with green hair, they might add \textit{"green"} or \textit{"green hair"} to the negative prompt. Others noted that SD could yield unforeseen pleasant outcomes beyond the intended prompt. P2 highlighted their openness to serendipitous discoveries while creating random non-playable characters. They employed SD to generate numerous renditions of vague concepts and refined the prompts by drawing inspiration from the preferred interpretations. Nonetheless, they noted the difficulty in refining their prompt to mirror the style of a prior generation.

\subsection{Managing Model Randomness is a Challenge}
\label{fs_large_collection}
Managing the randomness of images generated by SD from a single prompt was a difficulty mentioned by participants. P3 stated that the randomness could \textit{"drive me crazy because I might lose the essence of what I was looking for."} P6 noted that \textit{"when the prompt is too simple, [generated images] often jump between styles."} While P5 believed that \textit{"randomness is a big factor in the quality of the images."}, highlighting the significance of serendipity in text-to-image generation. To overcome these challenges, all participants mentioned generating large batches of images (30-50) to find images with the desired subject matter and style while avoiding anomalies. When conducting iterative experimentation, users with less powerful GPUs used small (2-10) batches to refine their prompts and only generated a large batch of images when they achieved satisfactory results. The participants' strategy to overcome model stochasticity echoes Liu et al.~\cite{liu2022opal}'s findings that generating more images gives users better outcomes and helps them understand AI capabilities.
\section{Design Goals}
\label{design_goals}
Guided by our formative interviews, we established three goals to guide the design and development of Promptify to facilitate the workflow of creating images with text-to-image models. 
Our system is designed to cater to novice users, who have been identified as the user group facing the most significant challenges when using these models.

\vspace{6pt}
\noindent
\textbf{D1. Assisting Users in Prompt Engineering.} 
  Effective prompt writing or prompt engineering is a task identified as a significant hurdle in our formative study and previous research~\cite{liu2022promptenineering, Saravia_Prompt_Engineering_Guide_2022}. Promptify aims to alleviate this challenge by assisting users in generating potential subject matters and suggesting effective keywords that can aid in achieving their creative objectives.

\vspace{6pt}

\noindent \textbf{D2. Supporting Iterative Experimentation of Text-to-Image Generation.} 
Promptify should support the iterative workflow that includes ideation of subject matters and styles, construction of effective prompts that reflect the user's creative intention, and refinement of initial prompts based on previously generated results, following the user's needs.

\vspace{6pt}

\noindent
\textbf{D3. Facilitating Examination and Organization of Generated Images}. 
Consistent with prior studies~\cite{liu2022opal}, participants in our formative study highlighted the significance of serendipity and the need to generate a substantial number of images to achieve optimal results. As such, Promptify aims to facilitate the practice of comparing, contrasting, and managing a large number of images for its users. 
\begin{figure*}[t]
  \centering
  \includegraphics[width=\linewidth]{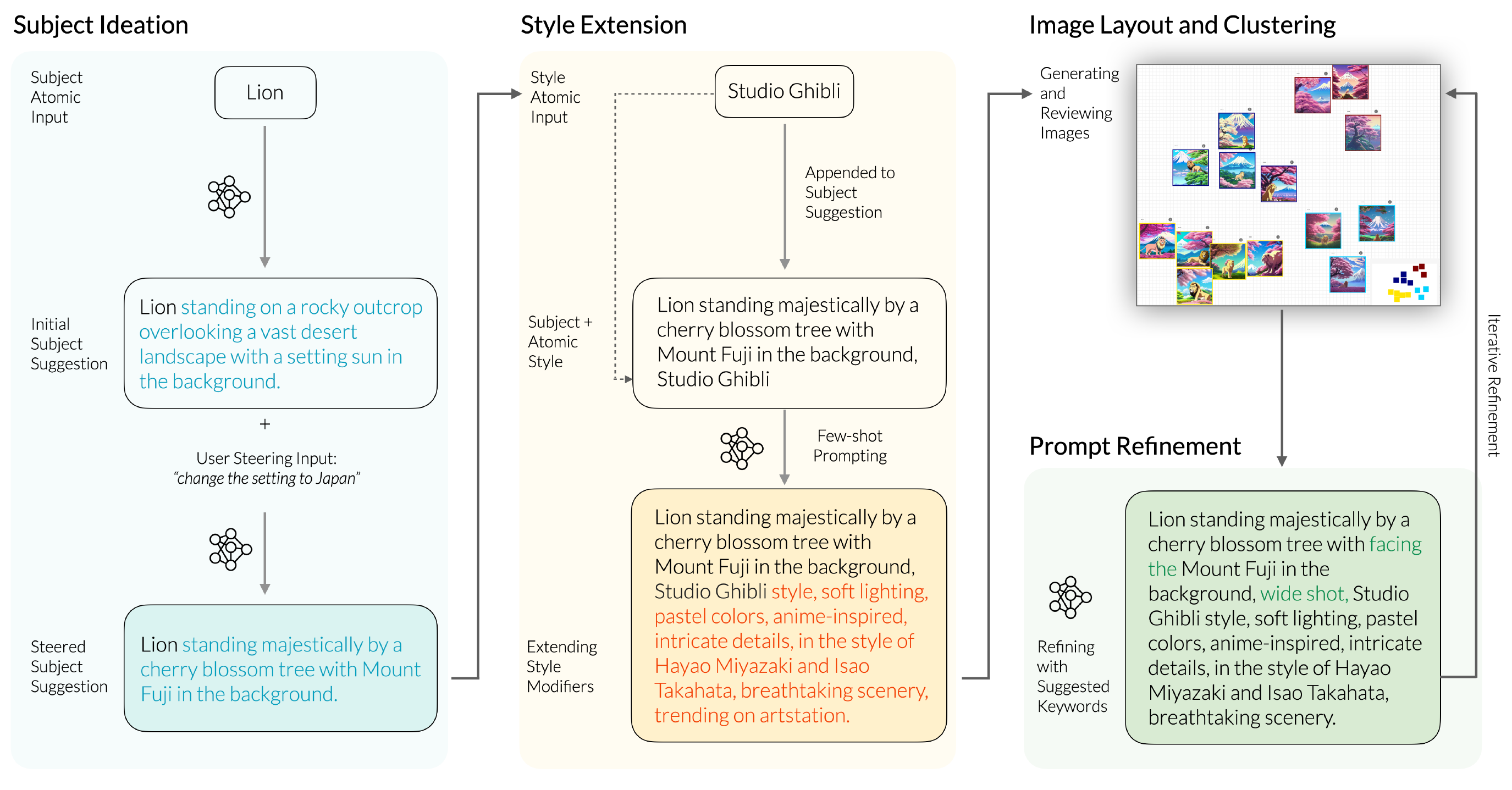}
  \caption{The user workflow with the Promptify system. Users start by inputting an atomic subject, e.g., \textit{Lion}, to obtain various subject ideas via the steerable subject ideation feature. Next, users input an atomic style such as \textit{"Studio Ghibli"} and use the style extension feature to extend the style portion of the prompt with their desired subject matter. Finally, users generate images, position and cluster them by similarity, effortlessly exposing themes between the images. To refine their work, they can search through the modifiers suggested per image and cluster via CLIP Interrogator to reinforce the desired features and discourage the unwanted ones for the next iteration. Users can skip Promptify's suggestion features at any time and manually write parts of the prompt if they choose to. The neural net icons indicate the component is assisted by the LLM.}
  \Description{}
  \label{fig:workflow}
\end{figure*}
\setstcolor{red}

\section{The Promptify System}
Promptify improves the workflow of text-to-image generation by offering three main features, including \textit{1) automatic prompt extension and suggestion, 2) image layout and clustering by similarity, and 3) automatic prompt refinement suggestions.} By integrating these features, Promptify establishes a feedback loop that enables users to generate high-quality images from initial prompts, examine how the model interprets their prompt, and receive suggestions to edit the initial prompts for enhancing output. Figure \ref{fig:workflow} illustrates the user workflow using Promptify. Next, we will introduce individual features and associate them with the design goals outlined in Section~\ref{design_goals} to underscore their design rationale.

\setstcolor{red}
\subsection{Automatic Prompt Suggestion}
\begin{figure}[h]
  \centering
\includegraphics[width=\linewidth]{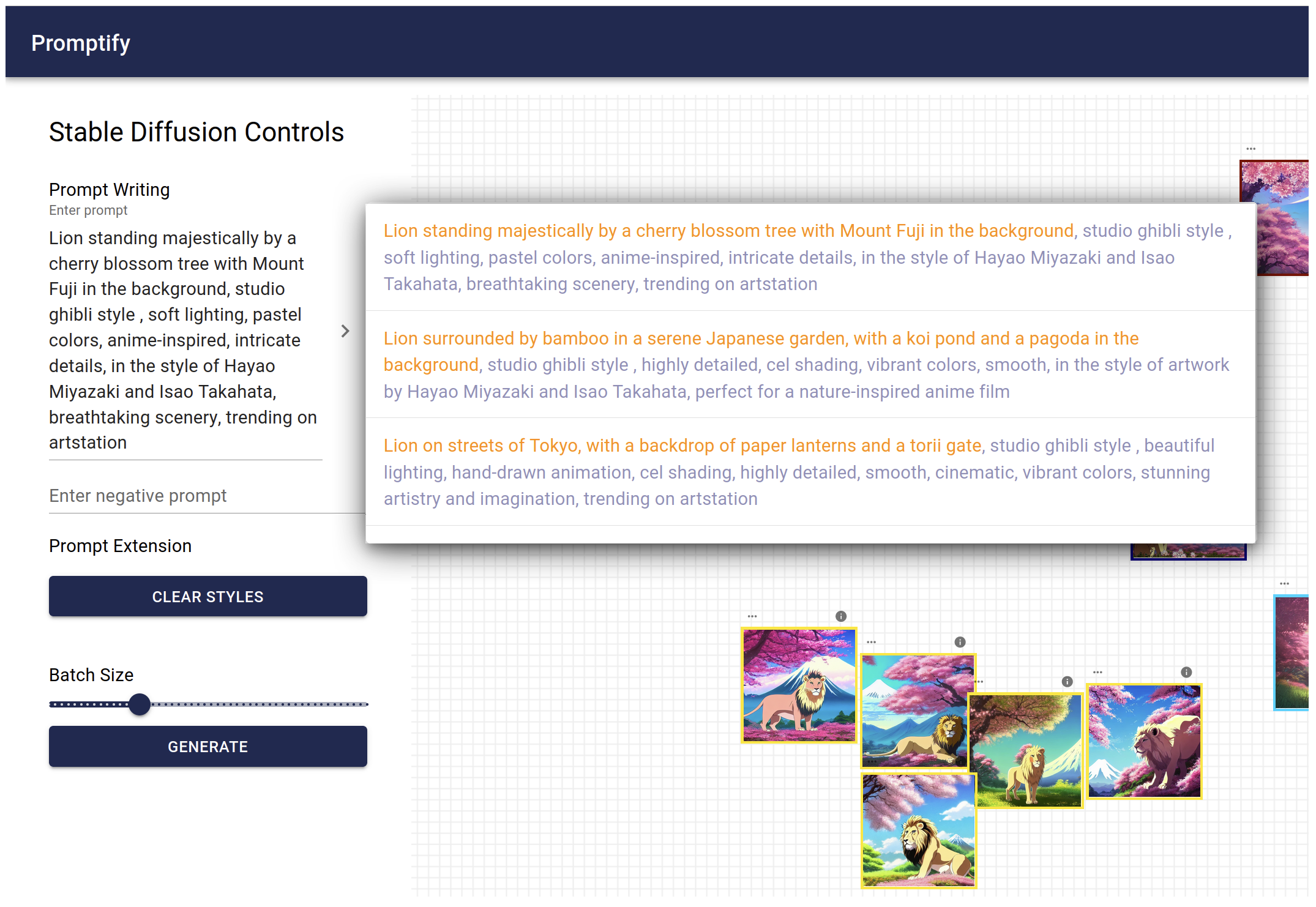}
  \caption{Promptify's prompt suggestion menu. The suggested subject phrases are colored orange, while the style modifiers are highlighted in violet. The first suggestion automatically propagates to the prompt writing text field. Users can further edit the suggested prompts.}
  \Description{}
  \label{fig:suggestion}
\end{figure}

A key challenge of using text-to-image models is constructing prompts so that the models can properly interpret the user's creative intention. To support prompt writing (\textbf{D1}), we developed a prompt suggestion engine, as shown in Figure~\ref{fig:suggestion}. The prompt suggestion process in our system is divided into  \textit{subject extension} and \textit{style extension}, each expanding upon an atomic input. An atomic input, in this context, refers to a basic description of either the subject or style. Note that users have the ability to manually edit any of the suggestions provided and can flexibly opt out of either extension as they see fit. 

\subsubsection{Subject Ideation} 
Users begin by entering an atomic subject, such as \textit{"Lion"}, in the prompt writing field. They can then choose to extend the subject by clicking on the "extend subject" buttons, and Promptify will add further details to the prompt. The system will present the user with a pop-up menu containing three options for suggested extensions. For example, the user may receive a suggestion \textit{"Lion standing on a rocky outcrop overlooking a vast desert landscape with a setting sun in the background."} To further refine the prompt, users can use the "steer the subject matter" text field to provide additional instructions to the system, as shown in Figure \ref{fig:teaser}-B. For example, they may request a winter-themed prompt by typing in \textit{"change the setting to Japan."}. The system will then generate a new prompt that incorporates the user's instruction, such as \textit{"Lion standing majestically by a cherry blossom tree with Mount Fuji in the background."}

\subsubsection{Style Extension} 
Once users have selected their desired subjects, they can enter a short style description that characterizes the high-level styles they wish to create in the "steer the style" field. Promptify will then apply this style to each suggested subject and extend the prompt with further relevant details. For instance, if the user types in \textit{"studio ghibli,"} Promptify extends the prompt to \textit{"{studio ghibli style, soft lighting, pastel colors, anime-inspired, intricate details, in the style of Hayao Miyazaki and Isao Takahata, breathtaking scenery, trending on artstation}"}. We adopt the {\{subject, style modifier 1, style modifier 2...\}} prompt format, which begins with subject keywords followed by multiple style modifiers \cite{wangDiffusionDBLargescalePrompt2022, oppenlaender2022taxonomy}. The style extension feature is designed to generate prompt suggestions resembling the language used by the SD online community to increase the likelihood of generating high-quality images (\textbf{D1}). 

\clearpage

\subsection{Image Layout and Clustering}
\begin{figure}[h]
  \centering
\includegraphics[width=\linewidth]{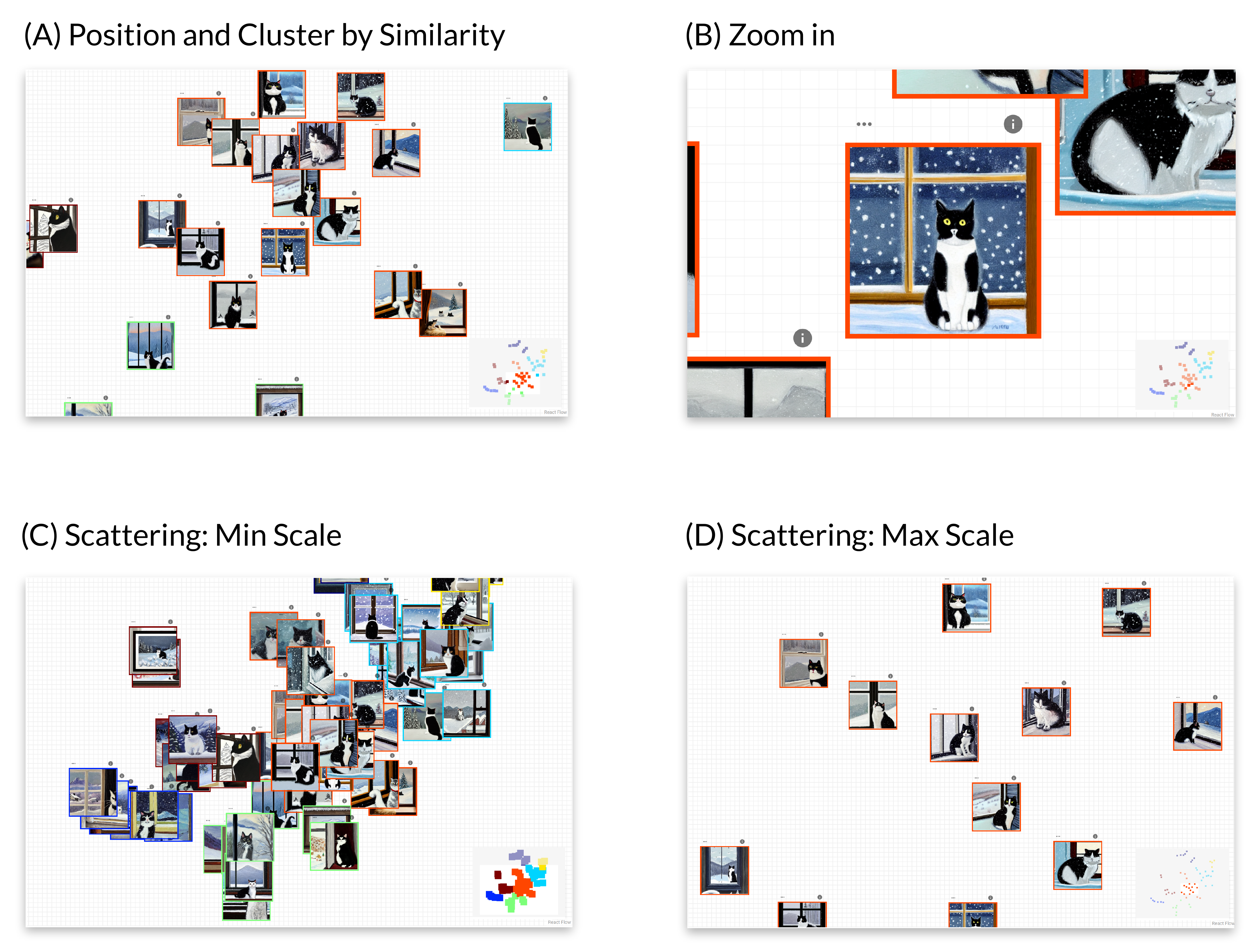}
  \caption{ Image browsing techniques supported by Promptify's interface. (A) Positioning and clustering by image similarity; the border colors of images indicate their cluster and are shown on the minimap. (B) Users can zoom in to inspect the generated images in their full size (512x512). (C) The minimum scale of image layout. Images are gathered and overlap with minimal white spaces. (D) The minimum scale of image layout. Images are widespread without any overlaps.} 
  \Description{}
  \label{fig:scattering}
\end{figure}

Upon finishing prompt writing, users can generate batches of images with Promptify. Our system supports the examination and management of the generated images (\textbf{D3}) by visualizing them on a 2D interactive zoomable canvas. As shown in Figure \ref{fig:scattering}, Promptify allows users to reorganize the generated images and cluster them based on similarity. These clusters are distinguished by distinct colors on the minimap, which serve as visual cues for users. Moreover, users can adjust the distance between images using the scale slider to avoid any disruptive overlap. They can compare and contrast different aspects of the images by dragging and rearranging them. The interface can also facilitate examination across multiple iterations (\textbf{D2}). If users like a particular image cluster, they can keep that cluster in mind for the next round of generation and attend to the new images that are positioned close to the cluster. The image layout and clustering feature helps users find desired images, ignore undesired ones, and discover thematic elements across their generated images. It is useful for further editing and refining the prompts.

\subsection{Suggesting Modifiers Based on Prior Generation}
\begin{figure}[h]
  \centering
\includegraphics[width=\linewidth]{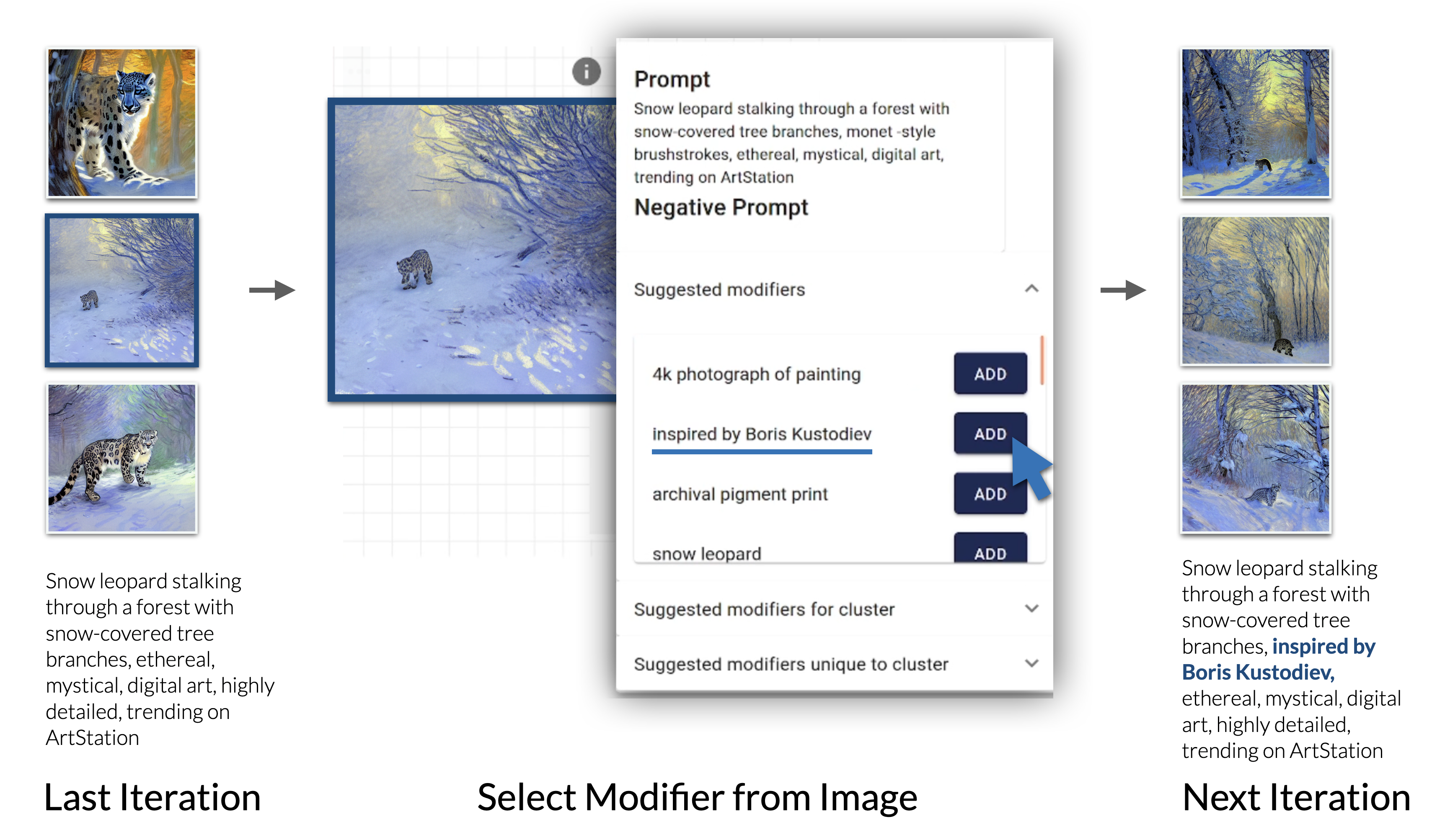}
  \caption{Example of our study participant using CLIP Interrogator's keyword suggestions to refine prompts to focus on specific styles. The user added modifiers suggested based on the image in the middle from the previous generation to create more images in that style.}
  \Description{}
\label{fig:modifier_menu}
\end{figure}

After generating images, users can enhance or avoid certain features in subsequent generations by editing the prompts. Promptify supports this practice by providing refinement modifiers, which are keywords or key phrases associated with the generated images and their corresponding cluster (\textbf{D2}). As depicted in Figure \ref{fig:modifier_menu}, users can access the modifiers by clicking on the information button of an image, which will display a pop-up menu with three collapsible sub-menus: one for modifiers associated with that particular image, one for modifiers associated with the cluster where the image belongs to, and one for modifiers unique to that cluster. Users can add any of these prompts by clicking the "Add" button, which triggers an automatic process to incorporate the modifier into their prompt. The suggested modifiers are intended to provide users with descriptive terms and phrases linked to particular images or themes within clusters, assisting them in adjusting their input prompts to reinforce preferred features or discourage undesirable ones.

\begin{figure*}[!t]
  \centering
  \includegraphics[width=\linewidth]{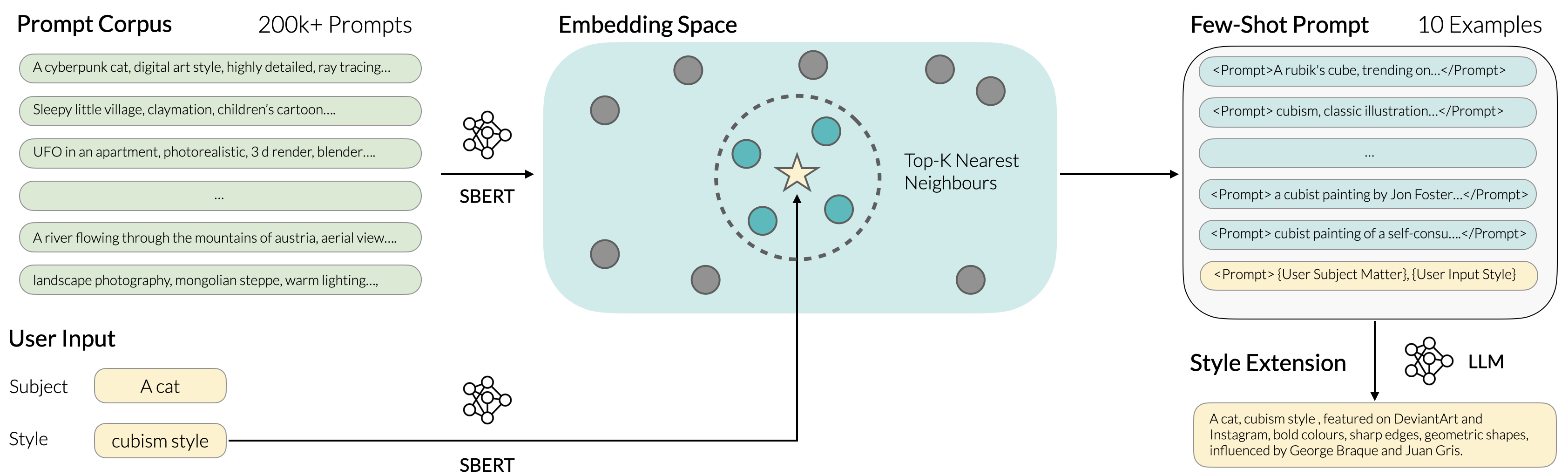}
  \caption{Illustration of our few-shot prompting pipeline for style extension. We pre-embedded the prompt corpus using sentenceBERT embedding. When the user inputs an atomic style to request an extension, Promptify embeds the input and uses it as a query to perform KNN search to retrieve semantically similar examples from the corpus. The retrieved examples and the user's input were then used to construct the few-shot prompt for the LLM to perform style extension.}
  \label{fig:knn}
  \Description{}
\end{figure*}

\section{BackEnd Implementation}

\subsection{Automatic Prompt Suggestion}
The prompt suggestion engine is implemented by prompting an LLM named GPT-3.5 (ChatGPT) developed by Open AI \cite{openaiapi}. The implementation of subject and style suggestions differs, with the former being accomplished via 0-shot instruction and the latter being achieved through few-shot prompting using community data. Our LLM approach is flexible as future zero shot or few shot prompts which prove to be useful for subject/style extension can be immediately integrated into our system. The prompts for each LLM-enabled feature can be found in the supplementary file. 

\subsubsection{Subject Ideation}
We employ a zero-shot prompt  to guide the LLM in generating suggestions for extending an atomic subject. Zero-shot refers to the absence of concrete examples provided to the model other than the instruction. We found the LLM is highly capable of ideating various subject suggestions when prompted only with the instruction.  Providing additional examples may introduce bias and narrow the space of possible suggestions, which is undesirable for our purpose that requires exploratory ideation. To contextualize text-to-image generation for the LLM, our prompt refers to DALL-E~\cite{ramesh2021zeroshot}, a well-known text-to-image model predating GPT-3.5's knowledge cutoff in September 2021. We instruct the model to generate three detailed prompt suggestions for a given atomic subject (e.g., "a cow"). To enable language-based steering of suggestions by users, we log the query and response that generated the initial set of suggestions. When users provide additional instructions requesting modifications, these instructions, along with the initial ones, are sent to the LLM to generate new suggestions.

\subsubsection{Style Extension}
We use few-shot prompting to teach the LLM how to extend the style of a prompt given the user's atomic style input. Compared to zero-shot learning, few-shot prompting uses a small set of task-specific examples to guide the LLM in generating outputs that follow similar patterns. This approach is advantageous for our goal of designing style extension features that suggest effective style modifiers in language similar to those used by the community~\cite{10.1145/1622176.1622214} \textbf{(D1)}. 

Once the user types in an atomic style and requests an extension, our system will construct a few-shot prompt consisting of 10 prompt examples relevant to the input style, sampled from a corpus containing community-shared SD prompts~\cite{wangDiffusionDBLargescalePrompt2022}.  The examples are sampled via a KNN algorithm based on sentence similarity as follows. We first constructed a prompt corpus by randomly sampling 500k prompts from the DiffusionDB \cite{wangDiffusionDBLargescalePrompt2022} dataset. Each prompt data came with a not safe for work (NSFW) score, and we filtered those with a score larger than 0.1. To guide the model to generate prompt suggestions in the \textit{{subject matter, style modifier 1, style modifier 2...}} format, we selected only prompts with at least 6 modifiers (phrases separated by commas). The filtering process resulted in the final corpus of 238584 prompts. We then used Sentence-BERT \cite{sbert} to embed all the prompts into sentence embeddings. When users enter an atomic style such as \textit{impressionism}, Promptify embeds the users' input phrases, e.g. the query, using the same embedding model. It then retrieves the 10 examples in the corpus with the highest cosine similarity to the query using the KNN algorithm (k=10). The effectiveness of the KNN approach has been demonstrated in prior research by Liu et al.  \cite{liu2021makes}, who found that selecting examples closer to the query in the embedding space consistently resulted in superior performance for few-shot example selection. As shown in Figure \ref{fig:knn}, the few-shot prompts consist of 10 queried examples. Following this, the user's present subject matter and atomic style are concatenated with the format and appended to the end of the few-shot examples. The final prompt is then inputted into the LLM to autocomplete the additional style modifiers, guided by the few-shot examples.
\subsection{Images Layout and Clustering}
The image layout and clustering feature is implemented based on CLIP embedding. CLIP ~\cite{clip} is a neural network trained on a variety of (image, text) pairs, with the objective of obtaining a joint multimodal embedding space of language and image. We chose CLIP embedding because it is also used to condition the denoising process in diffusion models like Stable Diffusion and DALL-E 2. To implement image layout and clustering, we first compute the CLIP image embeddings of all generated images on the canvas. Next, we utilize TSNE \cite{van2008visualizing} to reduce the dimensionality of the embeddings to 2D, which allows for easy visualization of the images. The reduced 2D embedding serve as the coordinates to plot the images, providing users with an intuitive view of their image collection. To prevent overlapping, we scale these coordinates so that the minimum distance between images is 128 pixels.  We cluster the images based on their position on the screen using affinity propagation \cite{affinity}, which does not require users to determine the number of clusters ahead of time. The scale slider, which allows users to spread images apart on the canvas to avoid overlaps, was implemented by multiplying the coordinates of the images by the slider value with a default value of 1 and a range between 0.5 and 3.

\subsection{Modifier Suggestion from Generated Images}

\subsubsection{CLIP Interrogator Prompt Modifier Suggestion}
We utilize CLIP Interrogator to tag the generated images with a list of prompt modifiers that users can choose from to refine their prompt from the last iteration. CLIP Interrogator is an open-source tool that leverages BLIP~\cite{blip} to generate a caption for the image and subsequently identifies a list of high-scoring keywords associated with the image. These keywords are selected from a corpus comprising approximately 100,000 words and phrases and 5,000+ artists' names \cite{hardprompts}. We selected CLIP Interrogator as it serves as an effective tool for generating prompts that can replicate an image's subject matter and style when used for SD\cite{hardprompts}. A limitation of the CLIP Interrogator is its reliance on a restricted set of artists and keywords, which may not encompass every possible artist.

\subsubsection{Integrating Selected Modifiers}
Promptify integrates the prompt modifiers recommended by the CLIP Interrogator when the users select them. This integration is achieved through the LLM by employing a few-shot prompt consisting of eight manually curated examples of keyword integration. We wrote the examples to guide the LLM to avoid redundancies while modifying the prompt. For instance, if the source prompt mentions \textit{"a brown cow,"} and the CLIP Interrogator modifier suggests \textit{"a cow with a red barn in the background,"} this few-shot prompt is crafted to integrate the two as \textit{"a brown cow with a red barn in the background."} Each few-shot example consists of three distinct parts: a source prompt obtained from diffusion DB, a modifier obtained from CLIP Interrogator, and the integration of the two. When the user selects a modifier, Promptify appends the user's current prompt and the modifier to the few-shot examples following the same format for LLM to auto-complete the combination. Among the eight few-shot examples, half of them involve CLIP Interrogator prompt modifiers related to the subject matter, and the other half is related to the style. We incorporated this auto-integration feature to alleviate the cognitive burden on users, who would otherwise be required to parse through redundancies and edit prompts themselves.

\subsection{{System Architecture}} 
We implemented the Promptify full-stack system with a Python Flask  and React Flow~\cite{reactflow}. The system runs on a Linux machine equipped with an Nvidia Titan RTX GPU, which takes ~4 seconds to generate a 512x512 image with SD. We access the OpenAI API \cite{openaiapi} to use the \texttt{GPT-3.5-Turbo} model for prompt suggestion and editing. We use the \texttt{all-MiniLM-L6-v2} model from the SentenceTransformers python library \cite{sbert} for sentence embedding. We access the CLIP embedding for image layout and clustering and the CLIP Interrogator using their respective open-sourced code on GitHub \cite{clip, clipinterrogator}. We use \texttt{scikit-learn} for affinity propagation.

\section{User Evaluation}
We conducted a user evaluation to evaluate Promptfiy's effectiveness in fulfilling our design goals to facilitate text-to-image workflow. The purpose of the study was to gain insight into how users perceive and interact with individual features of the system, and to identify any potential obstacles. The user evaluation was divided into 3 parts: comparative evaluation, free-form usage observation, and semi-structured interview.  For the comparative evaluation, participants used both Promptify and the commonly used open-source UI, Automatic1111. Next, they freely interacted with Promptify to create images. Lastly, we conducted semi-structured interviews to understand their experiences with Promptify. Both Promptify and Automatic1111 were installed on a desktop computer that had an NVIDIA Titan RTX GPU. We used the Stable-Diffusion-v1-5 checkpoint and set all hyperparameters to the same values, including 50 denoising steps, 7.5  CFG scale, and Euler A noise sampling.

\subsection{Participants}
We recruited 15 participants in total for the study and completed 14 user studies (Mean age=24.2, STD=1.7). During one of the studies, a participant who was a professional visual artist attempted to generate images in the style of Egon Schiele \footnote{Egon Schiele (Wikipedia): \url{https://en.wikipedia.org/wiki/Egon_Schiele}}. However, the images were repeatedly blocked by the NSFW content filter we had in place. The participant eventually decided to drop out of the study. Out of the 14 participants who completed the study, none of them had significant experience with SD. Nine participants reported having no prior experience with text-to-image models, while the remaining five had only used SD or DALLE a few times. Furthermore, none of the participants reported using text-to-image models regularly or using them for professional work. The study took approximately 1.5 hrs and the participants were compensated with 30 CAD.

\begin{figure*}[!t]
  \centering
\includegraphics[width=\linewidth]{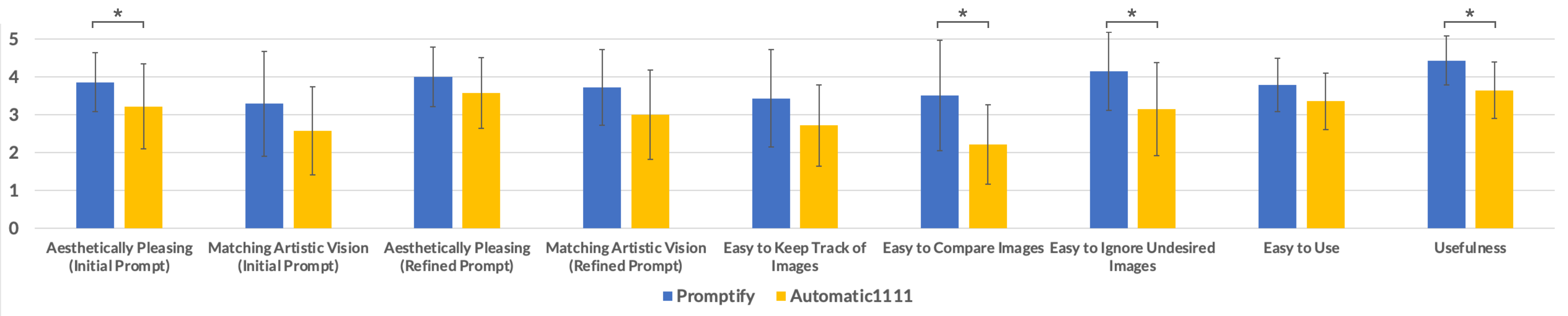}
  \caption{Results of the subjective ratings from the comparative evaluation. Promptify receives unanimously higher average scores than the baseline tool. Error bars represent standard deviation and the asterisks highlight results with p < 0.05. }
  \Description{}
    \label{fig:likert}
\end{figure*}

\subsection{Methodology}
We conducted the study in a quiet laboratory environment. Participants received a brief introduction to the study and the systems. After that, they tested each system and filled out the corresponding questionnaire. Semi-structured interviews were conducted after participants completed both the comparative evaluation and free-form usage observation. During the study, we encouraged participants to vocalize their thoughts and ask any questions they had.

\subsubsection{Comparative Evaluation}
\label{compare_methodology}
We employed a within-subject approach to compare the user experience between Promptify and Automatic1111 \cite{automatic1111}. We ask the participants to use the \texttt{txt2img} tab in Automatic1111 designed for text-to-image generation features. We counterbalanced the system order--half of the participants starts with Promptify and the other half with Automatic111. To assist participants in finding inspiration, we recommended they begin by selecting their favorite animal and an artist or visual art style they enjoy. These two choices would serve as starting subject matters and styles to be utilized in both interfaces. We acknowledge that this approach may limit creative freedom, yet it provided a common starting point for all participants to ensure a more controlled comparison. Participants were allowed to include additional details if desired. Below we outline the steps taken in each condition. These steps replicate the iterative process of prompt engineering that we discovered in the formative study, in a controlled manner.
\begin{steps}
  \item Write an initial prompt given their favorite animal and artistic style.
  \item Generate 10 images based on the initial prompt.
  \item Review 
  the generated images and choose the favorite and the least favorite images.
  \item Refine the prompt to reinforce desired features and discourage unwanted ones.
  \item Repeat the refinement process twice more.
  \item Generate 50 images with a favorite prompt
  \item Review the generated images and choose a favorite of the 50 images
\end{steps}

\subsubsection{Free Form Usage Observation}
Once the comparative evaluation was completed, we let the participants freely engage with the Promptify features to create images.

\subsubsection{Semi-structured interview}
Finally, We conducted semi-structured interviews lasting 10-20 minutes to understand users' experience with \textit{Promptify}, including the effectiveness of individual features and any difficulties they encountered.

\subsection{Subjective Ratings}
We developed subjective ratings to solicit user feedback after each condition in the comparative evaluation. All ratings are on a 5-point Likert scale (1--Strongly Disagree, 5--Strongly Agree), except for task load questions, which used the 7-point NASA-TLX scale. ~\cite{hart1988development}

\subsubsection*{Aesthetic Qualities}
We asked participants if they found their generated images aesthetically pleasing in both the initial and refined prompts. This question is associated with Steps 1-5.

\subsubsection*{Fulfilling Artistic Vision} 
We inquired if the participants' created images aligned with their artistic vision. The question was also associated with Steps 1-5, and asked for both the initial prompt and the refined prompts.

\subsubsection*{Tracking and Examining Generated Images} 
Participants rated their experience tracking and examine the large collection of images they generated. They rated the ease of keeping track of favorites, comparing images, and ignoring disliked ones. This relates to Steps 5-7.

\subsubsection*{Task Workload}
We asked the participants to provide subjective ratings for mental demand, effort, and frustration when using each system to generate images.

\subsubsection*{Usefulness and Ease of Use}
Participants assessed the usefulness and ease of use of each system. For Promptify, we additionally ask them to provide these ratings for individual features.



\subsection{Comparative Evaluation Results}
\begin{figure}[t]
  \centering
  \includegraphics[width=\linewidth]{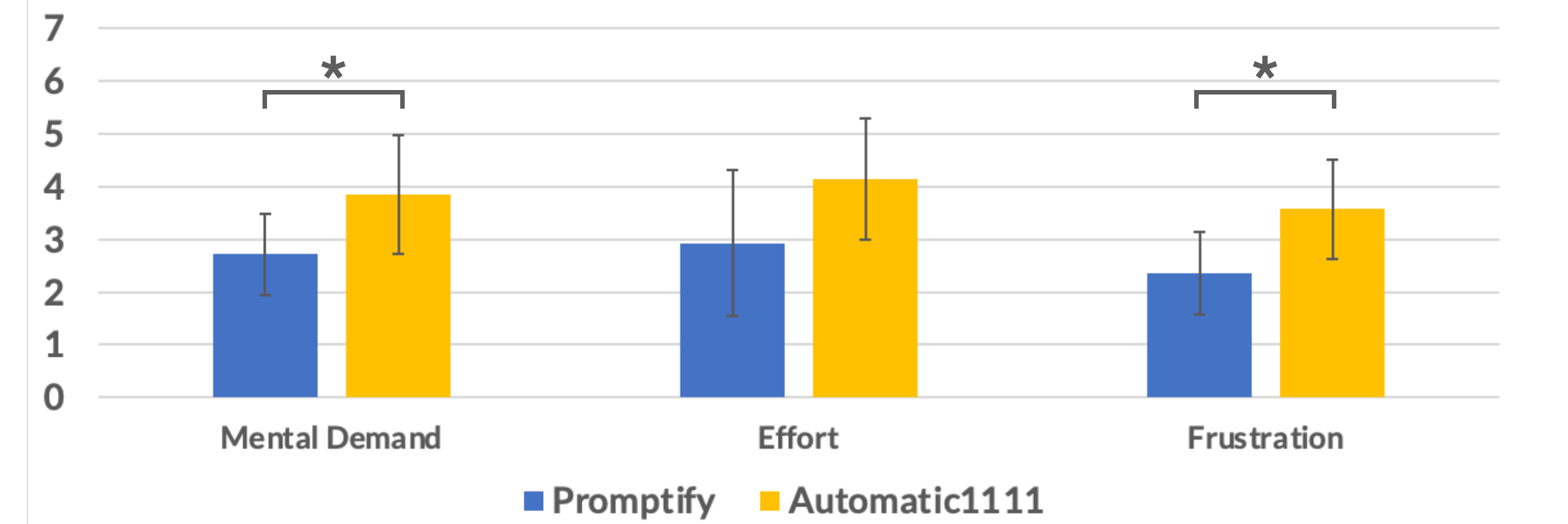}
  \caption{Results of the task load ratings: mental demand, effort, frustration. Error bars represent standard deviation and the asterisks highlight results with p < 0.05.}
  \Description{}
    \label{fig:nasa_tlx}
\end{figure}

As shown in Figure \ref{fig:likert} and \ref{fig:nasa_tlx}, Promptify received a higher average score unanimously across all ratings, demonstrating our system's effectiveness. Further analysis with a Wilcoxon Signed-Rank Test showed that our system is significantly more useful (Z = 11.0, p < .05) for supporting text-to-image generation workflow than the baseline tool. Moreover, users found the images generated with their initial prompts (Figure \ref{fig:first_batch}) significantly more aesthetically pleasing than the counterparts generated with the baseline (Z = 2.5, p < .05). Relevantly, the initial prompts written with Promptify were significantly longer (mean=34.43 words, STD=8.84) than those written with Automatic1111 (mean=8.6 words, STD=6.31) after stop-word removal using NLTK \cite{bird2009natural}. This demonstrates that Promptify's prompt suggestions empower users to write more detailed prompts to generate compelling visuals on their first attempt. Users had a significantly lower mental demand (Z = 9.5, p < .05) and less frustration (Z = 11.0, p < .05) when using Promptify. Participants also found using Promptfiy easier to compare and contrast images (Z = 13.0, p < .05) and to ignore images they do not like (Z = 13.5 p < .05) when searching for their favorite image in a large image collection.

\subsection{User Feedback on Promptify Features}
\begin{figure}[h]
  \centering  \includegraphics[width=\columnwidth]{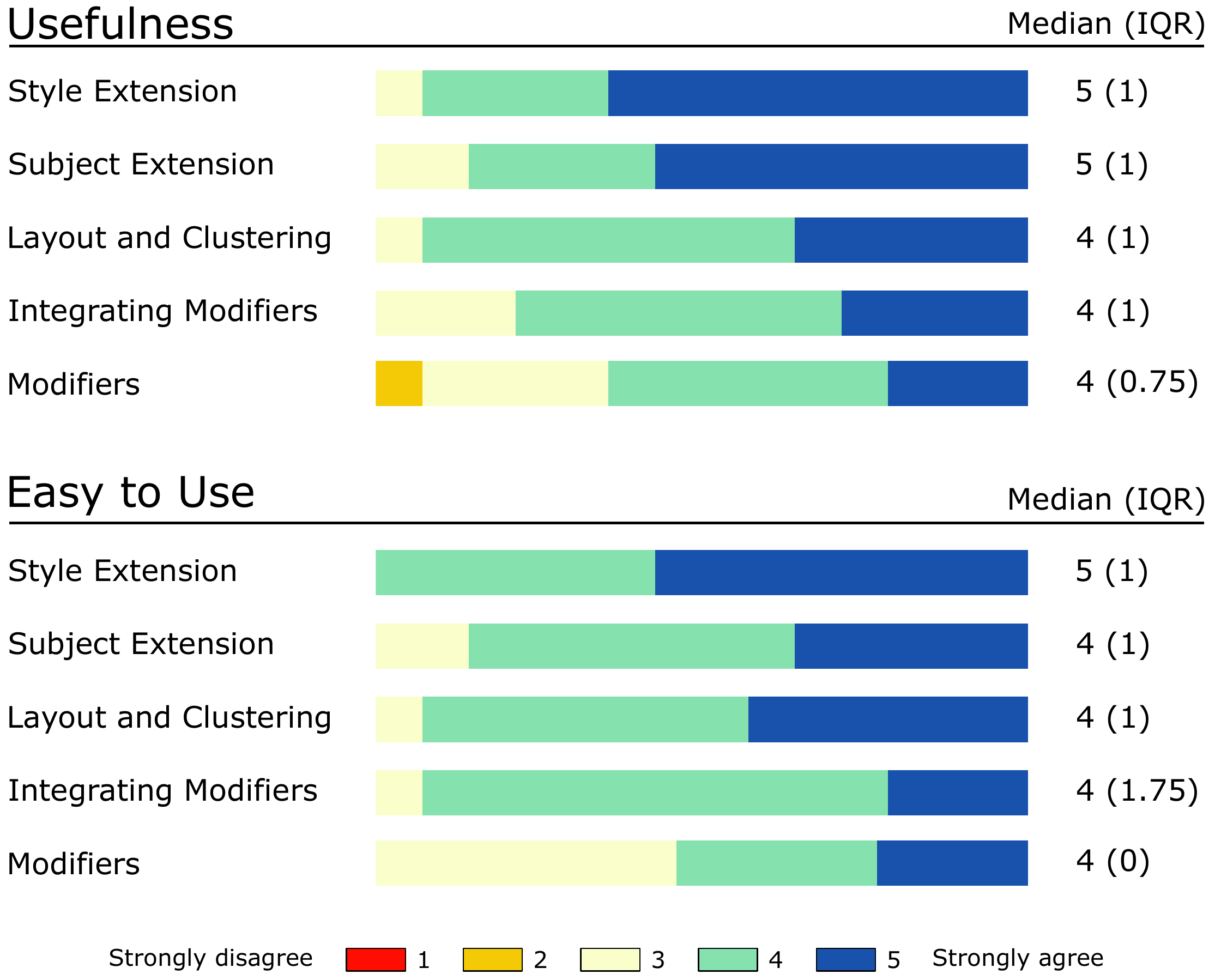}
  \caption{Results of subjective ratings of Promptify features showing participants largely agree that these features are easy to use and useful. Rows are sorted by mean usefulness score (Median, Inter-quartile Range). }
  \Description{}
  \label{fig:feature_feedback}
\end{figure}

We now discuss user feedback related to Promptify's features. This feedback is based on questionnaire and interview data, as well as observations taken during the comparative evaluation and free-form usage sessions. Subjective ratings of the Promptify features are illustrated in Figure \ref{fig:feature_feedback}.

\subsubsection{Subject Ideation}
\label{subject_results}
Participants found Promptify's subject suggestion engine useful and easy to use, with an average usefulness score of 4.43 (STD=0.76) and an ease of use rating of 4.21 (STD=0.70). Participants highlighted how subject ideation helped them fill in appropriate details of images that they would not have thought of on their own. P9 notes that \textit{"I found [Promptify] was less stressful in the beginning to get a first image. I felt like in [Automatic1111], maybe I didn't have the words to describe exactly what I had in my head.”} P1 tried to generate subject extensions for their abstract idea of "cloud fishing". She felt that \textit{"in this really abstract example, [Promptify] helped me refine my idea and do a really nice exploration with none of the frustration."} On the other hand, participants who already had a specific idea in mind found subject extension less useful. For example, P11, who wanted to generate an image of a sheep baking a pie, noted that since they already had a concrete target, the suggestions for subject matter were less relevant in the pursuit of creating that image.



\subsubsection{Style Extension}
Style extension is regarded as easy to use and is notably rated as our most useful feature, receiving a mean score of 4.57 (STD=0.65) for usefulness and a 4.56 (STD=0.51) for ease of use.  Participants highlighted how the style extension helped them achieve their desired style rapidly. For example, in the comparative study, P1 wanted to make an impressionist painting of a cat but felt that the images they generated with Automatic1111 \textit{"look too impressionistic--the colors are a bit too saturated"}. She further commented in the semi-structured interview \textit{“When I first started with Automatic1111 I would get stuck... I was stuck as to which prompts I should give to get the effect that I knew I wanted in my head"} and commented that the wealth of style suggestions made by Promptify made it much easier to achieve the style she wanted. Furthermore, P5 believed that style extension fills in specific key phrases that are specifically useful for enhancing SD-generated images, noting \textit{“I wouldn't have thought of adding all the ‘highly detailed’ or ‘art station trending’, ‘digital painting' tags."}. Other participants raised concerns that the style extension feature might suggest artists' names that they were unfamiliar with, which could impede their comprehension of the prompt. As P10 noted, \textit{"if you're an artist and you really know your styles well, you could probably use it much better."}

\subsubsection{Image Layout and Clustering}
Image layout and clustering received an average usefulness rating of 4.29 (STD=0.61) and an ease of use rating of 4.36 (STD=0.63). Participants reported that this feature helped them manage large batches of generated images by exposing specific themes and enabling them to compare images from a past prompt with a refined prompt. Some participants utilized the drag-and-drop feature to create groups of their favorite images and others used clustering and positioning by similarity features to identify themes within images. The simplest use case that users highlighted was the ability to ignore entire clusters that did not appeal to them. For example, P2 was generating graffiti images of cheetahs and noticed that some clusters were missing cheetahs entirely and commented that \textit{"very quickly you can almost rule out a group."} Other participants noted that the image layout helped them reflect on different choices they can make when writing future prompts. P4, who generated images of an NBA buzzer-beater, noted that the image layout and clustering feature was useful to get a good idea of the different types of images that could be generated. It also helped them decide \textit{"whether I want multiple players or I want people dunking or shooting, or how close-up, do I want a lot of confetti or not. It felt like it was really useful to be able to navigate those things quickly."} One participant (P10) was rather critical about the clustering feature and commented that \textit{"the clustering seemed to be on really superficial things that, if you were told that it's being clustered by it, you'd agree."} This suggests the potential limitation of clustering based on CLIP embedding and that similarity determined with CLIP embedding may not align with the perception of users.

\subsubsection{Prompt Modifiers from CLIP Interrogator}
The prompt modifiers provided by CLIP Interrogator were considered valuable by most participants with a usefulness score of 3.79 (STD=0.89), and easy to use, with a score of 3.86 (STD=0.86). The style modifiers helped participants reinforce a style that they like but are not able to describe themselves. For example, when P1 was generating images of impressionist cats, her favored images all had a suggested modifier \texttt{"by Tsuguharu Foujita"}. She used the keyword to reinforce the style of the city backdrop they liked for the following generations and commented \textit{that "selecting a few [modifiers] helped me lock down on a style that I liked."}. When P8 was generating images of snow leopards in the style of Monet, their favourite image in their first batch depicted trees drawn with a particular artistic flair. As a refinement, they chose to add \texttt{"by Boris Kustodiev"} to their prompt to generate more images in the style (Figure \ref{fig:modifier_menu}) and commented the new images were \textit{"more similar. The style of the trees, for sure."} We observed similar interactions happening for 7 of our participants including these examples, showcasing the feature's effectiveness in facilitating iterative prompt refinement to concentrate on the content desired by users. Some participants, however, reported uncertainty regarding the usage of modifiers as they were not familiar with names of the artists being displayed. For example, P12 explained, \textit{"The difficulty is that I didn't know many of the artists already. So it gave me names, which I could search up and I could research, but off the top of my head, it didn't give me immediate information."} The CLIP Interrogator modifiers provide helpful tools for users to narrow down their preferred style. However, some participants find it difficult to use them for prompt refinement without knowing the identity of the artist in question. This might have resulted in the lowest average score for usefulness and easy-to-use rating among all of Promptify's features.

\begin{figure*}[!t]
  \centering
  \includegraphics[width=\linewidth]{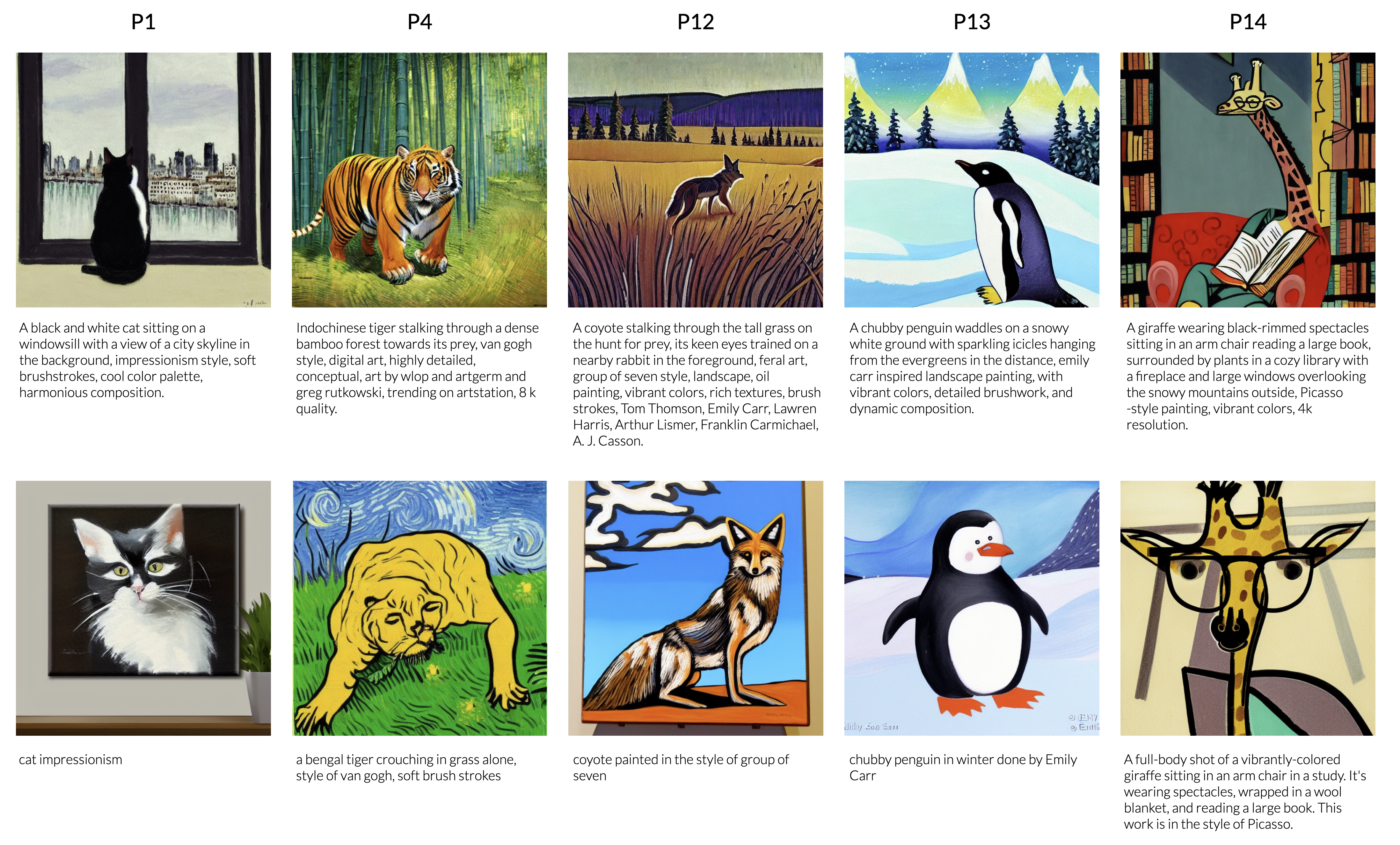}
  \caption{Example images from the first batch of images generated by study participants using Promptify (top row) and Automatic1111 (bottom row).}
  \Description{}
    \label{fig:first_batch}
\end{figure*}

\section{Discussion and Future Work}
\subsection{Facilitating Text-to-Image Workflow}
Promptify is designed to support the text-to-image workflow, consisting of prompt writing, generating and reviewing images, and refining initial prompts. 
The results of our user study indicate that Promptify is effective in suggesting prompts that lead to more aesthetically pleasing images compared to what users can generate on their own. The image layout and clustering UI is helpful in examining and comparing the generated images. 

While many participants found prompt modifiers suggested by CLIP Interrogator \cite{clipinterrogator} useful, we did not observe statistical significance to conclude that they enable better prompt refinement. The tool provides two types of keywords: 1) image captions for the subject matter and 2) artist names for styles. During experimentation, we found image captions tend to be less useful because they could be generic and overlap with the existing prompt's subject matter. Artist names were useful for generating art in specific styles but could be difficult for users who are not familiar with the artists. Future research can improve the suggestion of modifiers from generated images by utilizing captioning models that focus on specific regions or objects instead of the entire image to avoid generic captions. Additionally, incorporating stronger visual-language models such as GPT-4 \cite{gpt4} may generate additional style modifiers, augmenting the pre-curated list of artists in CLIP Interrogator. Promptify makes a comprehensive system contribution and can seamlessly incorporate advanced modules for each system component in the future.

\subsection{Automatic Prompt Suggestion with LLMs}
Promptify uses LLMs to assist with prompt writing for text-to-image generation. A unique aspect of our approach is the ability for users to guide the subject matter ideation through language instruction. Users found this feature helpful in diversifying their ideas and exploring novel subjects. Our style extension features use LLM few-shot prompting and community data to suggest styling keywords that mimic the community's language \cite{swales2014concept, 10.1145/1622176.1622214}. Although our current approach uses a specific format for prompts, we recognize that other formats can also be effective. However, we highlight the flexibility of our few-shot prompting approach as it can easily adapt to different prompt formats by providing several example prompts in the new format. An alternative approach to few-shot prompting  is finetuning an LLM \cite{magicprompt}. However, fine-tuning LLM is expensive and requires careful curation of the finetuning dataset. Our initial exploration found that finetuning often causes LLMs to forget their prior knowledge that can be useful for suggesting artists and styles (catastrophic forgetting \cite{driess2023palme}) and overfit on the data used for fine-tuning, e.g., always suggesting \texttt{"Greg Rutkowski"}, a digital artist popular in the SD community. Another future direction is to use techniques like Chain-of-Thought Prompting \cite{wei2023chainofthought} to guide LLMs in providing logical reasoning for suggested prompts. This can improve transparency and help users understand why certain keywords are suggested. For example, LLM can output explanations such as \textit{"Hayao Miyazaki is suggested to include for the style related to Ghibli Studio because he is a co-founder of the studio."} Finally, although the SD model we employ allows users to input negative prompts to guide the image generation away from undesired characteristics, our LLM engine currently does not suggest negative prompts. This area could be investigated in future research.

\subsection{Model Randomness and Effectiveness of Individual Keywords}
The process of generating images using a diffusion-based model is subject to randomness due to the use of random noise. This randomness can result in a wide variety of images generated from a single prompt. Moreover, not all intended keywords will always be reflected in the generated image. For example, a prompt requesting a cow on the grass may only produce images displaying the grass; a prompt for a Van Gogh chicken may result in a self-portrait of Vincent van Gogh instead of a chicken portrait in the artist's style (as intended by P6 in the user study). To mitigate the issue, Promptify organizes images semantically and allows users to remove undesired ones. As the length of the prompts increased, some participants faced difficulties in having specific details accurately depicted in the resulting images. This often occurred when the prompt suggestions contained multiple details, and users anticipated that all of them would be included in the resulting images. However, we found that the SD model tended to overlook phrases as the prompts became longer, resulting in images that lacked certain details specified in the prompts. For example, in Figure \ref{fig:first_batch}, P4-top is less Van Gogh like, and P12-top does not illustrate the rabbit described in the prompt. We hypothesize that this is the primary reason why Promptify did not significantly outperform Automatic1111 in terms of matching artistic vision. In addition, understanding how individual keywords and their position in a prompt impact the process of image generation remains a significant challenge. Our preliminary ablation analysis, which involves removing individual keywords from a prompt and determining sentence similarity with the original prompt using CLIP embedding, did not produce substantial evidence to inform the impact of individual keywords. It is crucial for future work to further explore methods to address model randomness and investigate the effect of individual phrases to assist users in creating effective prompts.

\subsection{Contrasting Promptify with Existing Tools}
The popularity of text-to-image models has led to the development of a range of supporting tools \cite{magicprompt, promptgen, automatic1111, auto1111_extensions}. We compared Promptify with Automatic1111, one of the most popular open-source tools, and showcased Promptify's substantial advantages. Some existing tools share similarities with a subset of Promptify's features. For instance, Magic Prompt \cite{magicprompt} and Promptgen \cite{promptgen} are both capable of suggesting prompts for text-to-image generation. However, Promptify provides a significantly more sophisticated prompt suggestion engine that enables separate suggestions for subject matter and style. Further, Promptify uniquely allows users to steer the subject ideation and style extension via an LLM. Moreover, Promptify introduces a novel image layout and clustering interface based on CLIP embedding, allowing users to browse and review images generated with SD.

Recent work in machine learning has started to develop advanced techniques to support image editing in text-to-image generation \cite{hertz2022prompt, zhang2023adding}. For example, Prompt-to-Prompt \cite{hertz2022prompt} leverages attention maps derived from diffusion models to enable layout-preserving image editing by modifying previous prompts. Our system focuses on a different but complementary goal, which is to aid novice users in identifying effective keywords that they might not have otherwise considered. 

In our current studies, we focused on experimenting with prompt writing and fixed all other vital parameters that can impact the text-to-image generation, such as random seed, noise scheduler, and CFG Scale. Additionally, there exist other variants of the SD model that allow for various types of user input to guide the image generation, such as img2img \cite{rombach2021highresolution}, inpainting \cite{rombach2021highresolution}, and ControlNet \cite{zhang2023adding}. Future work can integrate these parameters and models into Promptify, thereby enabling a broader range of capabilities for image generation.

\section{Conclusion}
We have presented Promptify, an interactive system supporting users to interactively explore and create text prompts for text-to-image generation. Promptify leverages an LLM to provide prompt suggestions separately for the subject matter and style. The subject matter suggestions are steerable by users via natural language instruction. The style keywords are suggested by few-shot prompting the LLM with text-to-image prompt examples from the online community. Promptify provides a layout and clustering interface that allows users to organize, review, and compare images efficiently. Promptify further suggests keywords associated with each generated image; users can use them to refine their initial prompts. Our 14-participant user study showed that Promptify is significantly more useful than a widely-used baseline tool for text-to-image generation, allowing users to create more visually appealing images on their first attempt while requiring significantly less cognitive loads.


\bibliographystyle{ACM-Reference-Format}
\bibliography{references}

\end{document}